\documentclass[12pt]{article}
\usepackage{amsmath}
\usepackage{amsfonts}
\usepackage{mathrsfs}
\usepackage{mathdots}
\usepackage{mathtools}
\usepackage{graphicx}
\usepackage{cleveref}
\graphicspath{ {./} }
\usepackage[backend=bibtex,
style=authoryear,
sorting=nty
]{biblatex}
\usepackage[titletoc,title]{appendix}
\usepackage{chngcntr}
\usepackage{apptools}

\newtheorem{theorem}{Theorem}

\newtheorem{proposition}{Proposition}
\newcommand{\brtheta}{\bar{\theta}}
\newcommand{\bbrtheta}{\boldsymbol{\bar{\theta}}}
\newcommand{\brlambda}{\boldsymbol{\bar{\lambda}}}
\newcommand{\brSigma}{\boldsymbol{\bar{\Sigma}}}
\newcommand{\brTheta}{\boldsymbol{\bar{\Theta}}}
\newcommand{\bXi}{\boldsymbol{\Xi}}

\newcommand{\bX}{\boldsymbol{X}}
\newcommand{\bA}{\boldsymbol{A}}

\newcommand{\bK}{\boldsymbol{K}}
\newcommand{\bone}{\boldsymbol{1}}
\newcommand{\bgamma}{\boldsymbol{\gamma}}
\newcommand{\brgamma}{\bar{\gamma}}
\newcommand{\bZ}{\boldsymbol{Z}}
\newcommand{\bhX}{\hat{\boldsymbol{X}}}

\newcommand{\bep}{\boldsymbol{\epsilon}}

\newcommand{\bhep}{\hat{\boldsymbol{\epsilon}}}

\newcommand{\btheta}{\boldsymbol{\theta}}
\newcommand{\bTheta}{\boldsymbol{\Theta}}
\newcommand{\blambda}{\boldsymbol{\lambda}}
\newcommand{\bPhi}{\boldsymbol{\Phi}}

\newcommand{\bXtheta}{\boldsymbol{X}_{\theta}}

\newcommand{\lag}{\textsc{lag}}
\newcommand{\SBT}{\textsc{SBT}}
\newcommand{\NLLK}{\textsc{NLLK}}

\newcommand{\mLb}{\bar{\mathscr{L}}}
\newcommand{\bOmega}{\boldsymbol{\Omega}}
\newcommand{\bSigma}{\boldsymbol{\Sigma}}
\newcommand{\bhTheta}{\boldsymbol{\hat{\Theta}}}
\newcommand{\ThetaSTq}{\boldsymbol{\Theta_{*,T-q}}}
\newcommand{\ThetaT}{\boldsymbol{\Theta_T}}

\DeclareMathOperator{\Tr}{Tr}
\DeclareMathOperator{\Cov}{Cov}

\title{Vector Autoregressive Moving Average Model with Scalar Moving Average}
\author{Du Nguyen \\ nguyendu@post.harvard.edu}
\begin{document}
\maketitle
\abstract{We show Vector Autoregressive Moving Average models with scalar Moving Average components could be estimated by generalized least square (GLS) for each fixed moving average polynomial. The conditional variance of the GLS model is 
the concentrated covariant matrix of the moving average process. Under GLS the likelihood function of these models has similar format to their VAR counterparts. Maximum likelihood estimate can be done by optimizing with gradient over the moving average parameters. These models are inexpensive generalizations of Vector Autoregressive models. We discuss a relationship between this result and the Borodin-Okounkov formula in operator theory.}
\section{Introduction}
Let $\bbrtheta(L) = \brtheta_0 + \brtheta_1 L + \cdots + \brtheta_qL^q$ be a polynomial matrix of size $s\times s$. Let $\Xi$ be a positive definite symmetric matrix of size $s\times s$.
Let $T > 0$. Consider
\begin{equation} \brgamma_l = \left\{ \begin{array}{l l} (\brtheta_0\bXi\brtheta_l^{\prime} +\brtheta_1\bXi\brtheta_{l+1}^{\prime}+\brtheta_2\bXi\brtheta_{l+2}^{\prime}\cdots + \brtheta_{q-l}\bXi\brtheta_q^{\prime}) & \text{for }l=0,1,\cdots,q \\
0 & \text{for }l > q
\end{array} \right.
\end{equation}
For each block matrix sequence $\bA = [A_0, A_1,\cdots,A_T]$ the associated Symmetric Block Toeplitz matrix (SBT) is given by:

\begin{equation}
\SBT(\bA) =\begin{pmatrix} A_0 & A_1 & A_2 & \cdots & \cdots & A_{T-1} & A_T \\
A_1^{\prime} & A_0 & A_1 & \cdots & \cdots & A_{T-2} & A_{T-1} \\
\vdots&\ddots &\ddots  &\ddots &\ddots &\ddots &\vdots \\
\vdots&\ddots &\ddots  &\ddots &\ddots &\ddots &\vdots \\
\vdots&\ddots &\ddots  &\ddots &\ddots &\ddots &\vdots \\
\vdots&\ddots &\ddots  &\ddots &\ddots &\ddots &\vdots \\
A_{T-1}^{\prime} & A_{T-2}^{\prime} & \cdots &\cdots & A_1^{\prime} & A_0 &A_1 \\
A_T^{\prime} & A_{T-1}^{\prime} & A_{T-2}^{\prime} & \cdots &\cdots & A_1^{\prime} & A_0
\end{pmatrix}
\end{equation}
In particular if $\bA = [\brgamma_0 \brgamma_1\cdots \brgamma_q, 0\cdots 0]$ then $\bA$ is
\begin{equation}
\brSigma_T =\begin{pmatrix} \brgamma_0 & \brgamma_1 & \brgamma_2 &\cdots &\brgamma_q& 0 & \cdots & 0 \\
\brgamma_1^{\prime} & \brgamma_0 & \brgamma_1 & \brgamma_2 & \cdots & \brgamma_q &\cdots & 0 \\
\vdots&\ddots &\ddots &\ddots &\ddots &\ddots &\ddots &\vdots \\
\vdots&\ddots &\ddots &\ddots &\ddots &\ddots &\ddots &\vdots \\
\vdots&\ddots &\ddots &\ddots &\ddots &\ddots &\ddots &\vdots \\
\vdots&\ddots &\ddots &\ddots &\ddots &\ddots &\ddots &\vdots \\

0 &\cdots & 0 &\brgamma_q^{\prime} &\cdots & \brgamma_1^{\prime} & \brgamma_0 &\brgamma_1 \\
0 &\cdots & 0 & 0 &\brgamma_q^{\prime} &\cdots & \brgamma_1^{\prime} & \brgamma_0
\end{pmatrix}
\end{equation}
which is a concentrated covariant matrix of the $VMA(q)$ process associated to $\brtheta$ where the covariant matrix of the innovation process is given by $\bXi$.

Let $\brTheta_T$ be the block matrix of size $Ts\times Ts$; $\brTheta_*$ be the matrix of size $qs\times qs$; $\brTheta_{*;T-q}$ of size $Ts\times qs$; $\brlambda$ be the matrix of size $Ts\times qs$ and $\bK(\brtheta, T)$ be the $Ts\times Ts$ matrix defined below:
$$\brTheta_T = 
\begin{pmatrix}
 \brtheta_0 & 0 & \cdots & 0 & 0 & 0\\
\brtheta_1 & \brtheta_0 & 0 & \cdots &  0 & 0 \\
\vdots& \vdots& \vdots & \vdots & \vdots & \vdots\\
\brtheta_{q-1} & \brtheta_{q-2} & \cdots & \cdots & 0 & 0\\
\brtheta_q & \brtheta_{q-1} & \brtheta_{q-2} & \cdots & 0 & 0\\ 
0 & \brtheta_q & \brtheta_{q-1} &\cdots & 0 & 0\\
\vdots & \vdots & \vdots & \vdots & \vdots & \vdots \\
0 & 0& 0 & \cdots & \brtheta_1 & \brtheta_0
\end{pmatrix}$$
$$\brTheta_* = 
\begin{pmatrix}
\brtheta_{q}& \brtheta_{q-1}&\cdots &\cdots & \cdots & \brtheta_1\\
0 &  \brtheta_{q}& \brtheta_{q-1} &\cdots &\cdots& \brtheta_2\\
0 & 0 & \brtheta_{q} & \brtheta_{q-1} & \cdots& \brtheta_3 \\
\vdots &\vdots & \vdots &\vdots & \vdots\\
0 & 0  & \cdots  &\cdots & 0 &\brtheta_q\\
\end{pmatrix}$$
\[
\brTheta_{*;T-q} = \begin{pmatrix}\brTheta_* \\ 0_{(T-q)s, qs} \end{pmatrix} 
\]
$$\brlambda = \brTheta_T^{-1}\brTheta_{*;T-q}$$
\begin{equation}
\label{eq:eqK}
\begin{aligned}
\bK(\bbrtheta, T) & = I_{T}\otimes\bXi^{-1} -(I_T\otimes \bXi^{-1})\brlambda [\brlambda^{\prime}(I_T\otimes\bXi^{-1}) \brlambda+I_{q}\otimes \bXi^{-1}]\brlambda^{\prime}(I_T\otimes \bXi^{-1})\\
& = (I_T\otimes \bXi+\brlambda(I_q\otimes \bXi)\brlambda^{\prime})^{-1} 
\end{aligned}
\end{equation}
$$\bSigma_T^{f} = \textsc{SBT}[\brgamma_0, \cdots \brgamma_q, 0, \cdots\brgamma_q^{\prime}\cdots\brgamma_0^{\prime}]$$
The second equality in \cref{eq:eqK} is an application of Woodbury matrix identity. The result of section 3.4 of \parencite{PhadkeKedem} is essentially the following proposition:
\begin{proposition}
\label{prop:gammaf}
\begin{equation}
\brSigma_T^{f} = \brTheta_T (I_T\otimes \bXi) \brTheta_T^{\prime}
\end{equation}
\begin{equation}
\label{eq:gamma_expand}
\brSigma_T = \Sigma_T^f + \brTheta_{*;T-q} (I_q\otimes \bXi) \brTheta_{*;T-q}^{\prime}
\end{equation}
\begin{equation}
\begin{aligned}
\brSigma_T^{-1}=&(\brSigma_T^{f})^{-1} - \\
& (\brSigma_T^{f})^{-1} \brTheta_{*;T-q}[I_q\otimes \bXi^{-1}+ \brTheta_{*;T-q}^{\prime}(\brSigma_T^{f})^{-1}\brTheta_{*;T-q}
]\bTheta_{*;T-q}^{\prime}(\brSigma_T^{f})^{-1} \\
= & \brTheta_T^{\prime-1} \bK(\bbrtheta, T) \brTheta_T
\end{aligned}
\end{equation}
\end{proposition}
This allows an efficient calculation $\brSigma_T^{-1}$ in the $VMA$ model. Consider the scalar case:
$$\btheta(L) = 1 + \theta_1 L + \cdots + \theta_qL^q$$
with $\theta_1,\cdots \theta_q$ are scalars. We will preserve the variable names but drop the bars on the variables in the scalar case, and will assume $\bXi = (1)$. In this case:
$$\bK(\btheta, T) = I_{T} -\blambda [\blambda^{\prime}\blambda+I_{q}]\blambda^{\prime}  = (I_T+\blambda\blambda^{\prime})^{-1} $$

Let
\[\bPhi(L) = I_k - \Phi_1 L - \cdots - \Phi^p L^p\]
Consider the $k$-dimension VARMA model with scalar:
\begin{equation}
\label{eq:VARsMA}
X_t = \mu +  X_{t-1} \Phi_1+ X_{t-2} \Phi_2+\cdots+ X_{t-p} \Phi_p +\epsilon_t + \theta_1 \epsilon_{t-1}+\cdots+\theta_q \epsilon_{t-q}
\end{equation}

The main result of this paper is the following:
\begin{theorem}
We have the following matrix identity:
\begin{equation}
\bSigma_T = \bTheta_T \bK(\btheta, T)^{-1}\bTheta_T^{\prime}
\label{eq:sigmatheta2}
\end{equation}  
The conditional log-likelihood function of the model in \cref{eq:VARsMA} conditioned on the first $p$ observations ($X_1,\cdots, X_p$) of the $T+p$ observations $X_1,\cdots,X_p,X_{p+1},\cdots X_{T+p}$ with $\theta_1,\cdots \theta_q$ are scalars is given by the formula
\begin{multline}
\mathscr{L}(\btheta, \mu, \bPhi,\bOmega,X_{p+1}\cdots X_{T+p}|X_1\cdots X_p) = -\frac{Tk}{2}\log(2\pi) -\frac{T}{2}\log(\det(\bOmega)) \\- {k/2} \log(\det(\blambda^{\prime}\blambda+I_q)) 
-\frac{1}{2}\Tr(\bZ^{\prime}\bTheta_T^{-1\prime}\bK(\btheta,T)\bTheta_T^{-1} \bZ\bOmega^{-1}))
\label{eq:llk1}
\end{multline}
where $\theta_0=1$,
$\btheta =(\theta_1,\cdots,\theta_q)$,
$\Phi=(\Phi_1,\cdots,\Phi_p)$,
$\bOmega$ is the covariance matrix of the $i.i.d.$ Gaussian random variables $\epsilon_i$'s. Here:
\[ \bZ =  \bX-\mu - L \bX \Phi_1-...- L^p \bX \Phi_p\]

\[\bX= \begin{pmatrix} X_{p+1}\\ \cdots \\ X_{T+p}\end{pmatrix}\]
of size $T\times k$.
\[L^i\bX = \begin{pmatrix}X_{p-i+1}\\ \cdots \\ X_{T+p-i}\end{pmatrix}\]

The optimal value is obtained at
\begin{equation}
\label{eq:opt_coeff}
\begin{pmatrix}\mu \\ \Phi_1\\ \Phi_2\\ \vdots \\ \Phi_p \end{pmatrix}_{opt} =  (\bX_{\btheta,\lag}^{\prime}\bK\bX_{\theta,\lag})^{-1} \bX_{\btheta,\lag}^{\prime}\bK\bXtheta
\end{equation}
where:
\begin{equation}
\bXtheta = \ThetaT^{-1}\bX
\end{equation}
\[ \bX_{\btheta,\lag}= \begin{bmatrix}\bTheta_T^{-1}\bone & \bTheta_T^{-1}L\bX &\cdots &\bTheta_T^{-1}L^p\bX
\end{bmatrix}
\]
and
\begin{equation}
\bOmega_{opt}(\btheta) = \frac{1}{T}
[\bXtheta^{\prime} \bK\bXtheta - 
 \bXtheta^{\prime} \bK \bX_{\btheta,\lag} (\bX_{\btheta,\lag}^{\prime}\bK\bX_{\btheta,\lag})^{-1} \bX_{\btheta,\lag}^{\prime}\bK\bXtheta]
\end{equation}
$\bOmega_{opt}$ is positive semi-definite regardless of sample values of $X$ and choice of $\btheta$. With these values of $\Phi_{opt}$ and $\bOmega_{opt}$, \cref{eq:llk1} is reduced to:
\begin{multline}
\mLb(\btheta,X_{p+1}\cdots X_{T+p} | X_1\cdots X_p) = -\frac{Tk}{2}\log(2\pi)-\frac{T}{2}\log(\det(\bOmega_{opt}(\btheta)))-\\
\frac{k}{2}\log(\det(\blambda^{\prime}\blambda+I_q))-\frac{Tk}{2}
\label{eq:llk2}
\end{multline}

Also we have:
\begin{equation}
\det(\blambda^{\prime}\blambda +I_q) = \det(\bSigma_T)=\frac{1}{\det(\bK(\btheta, T))}
\label{eq:sigmatheta3}
\end{equation}
\end{theorem}
We abbreviate VARMA models with scalar moving average components as VARsMA. We note this likelihood function is conditional only on the $p$ observations of $X$, and not on the initial error estimates $\epsilon$ in contrast with the typical conditional sum of squares (CSS) approach. In particular, for VMA models with scalar $\btheta$, the formula gives an exact likelihood formula. For scalar MA models, the formula for the likelihood function in term of $\bSigma_T$ is the same as those found in standard text books, e.g. \parencite{BJ, BrockwellDavis, Hamilton}. We first tried to compute VARsMA likelihood function conditioning on the pre-sample values of $\epsilon$ then integrating over them and rediscovered \cref{prop:gammaf} for the scalar case. The determinant of $\bSigma_T$ in \cref{eq:sigmatheta3} is one studied in the strong Szeg\"o limit theorem and the Borodin-Okounkov's determinant formula \parencite{GeCase,BO,BasorWidom} in the theory of Toeplitz operators, which we will discuss in section \cref{sec:history}.

Likelihood function for VARMA model is generally computed via Kalman filter \parencite{HarveyPhillips}. We note it could also be computed via tensor representation \parencite{NichollsHall}. Our formula is a simple generalization of the VAR case. It could find applications as an inexpensive enhancement to VAR. We note the approach of using generalized least squares for AR parameters has appeared in \parencite{HillmerTiao, ChibGreenberg, OttoBellBurman}. Our contribution is the observation that when $\btheta(L)$ is scalar, the moving average and the autoregressive polynomials commute, therefore we can apply generalize least square. This does not hold for VARMA in general. Also, while gradient method for VARMA is in general tedious, it is rather straight forward to compute gradient in this case. Combining GLS with \cref{prop:gammaf} gives us an efficient algorithm to estimate the parameters. We have implemented the algorithm in a python package, as well as in R.

Using $\bSigma_T$, \cref{eq:opt_coeff} could be rewritten as:
\begin{equation}
(\bX_{\lag} \Sigma_T^{-1}\bX_{\lag})^{-1} \bX_{\lag}^{\prime}\Sigma_T^{-1}\bX
\label{eq:ylk}
\end{equation}
with 
\[ \bX_{\lag}= \begin{bmatrix} \bone & L\bX &\cdots &L^p\bX\end{bmatrix}
\]

\begin{equation}
\Omega_{opt}(\theta) = \frac{1}{T}
[\bX^{\prime} \Sigma_T^{-1}\bX -
 \bX^{\prime} \Sigma_T^{-1} \bX_{\lag} (\bX_{\lag}^{\prime}\Sigma_T^{-1}\bX_{\lag})^{-1} \bX_{\lag}^{\prime}\Sigma_T^{-1} \bX]
\end{equation}

Any rational matrix transfer function could be brought to a form where the denominator is scalar. However after that transformation the numerator will have extra degrees, and generally not of full rank. As mentioned, we would like to consider our approach as an inexpensive enhancement to VAR. If we attempt to use pure VAR to model a process which has a slow decay moving average component, the VAR model would need to be of high order. If by adding one MA component, we can reduce the total degree $p$ of the numerator VAR process, which in general requires $k^2$ coefficients per extra degree, a VARMA with scalar MA model would be competitive in term of information efficiency. This model could be considered as a smoothing then regressing model where we have an efficient method to search for smoothing parameters. The likelihood formula is valid for any sample size, with no restriction on location of roots of $\btheta$. However for invertible $\btheta(L)$, the terms of $\bTheta_T^{-1}$ converges as $T$ increase. As $\bSigma_T$ is invariant under root inversion of $\btheta$, we can restrict our search to invertible moving average component.

We will use the same symbol $\btheta$ to denote both the polynomial $\btheta(L)$ and the vector $(\theta_1,\cdots \theta_q)$ of its non constant coefficients. Since we always refer to the polynomial with a variable, this will not cause confusion.

\section{Proof of the theorem}
Let $Z_t$ be the time series defined by:
\begin{equation} Z_t = X_t- \bone\mu- X_{t-1}\Phi_1  -\cdots- X_{t-p}\Phi_p = \epsilon_t + \theta_1 \epsilon_{t-1}+\cdots+\theta_q \epsilon_{t-q}
\label{eq:zeq}
\end{equation}
Assuming we have $n=T+p$ samples $X_{1},\cdots,X_p,X_{p+1},\cdots,X_{T+p}$ considered as rows of a matrix 
\[\bhX= \begin{pmatrix} X_1\\ \cdots \\ X_{T+p}\end{pmatrix}\]
of size $(T+p)\times k$. Let 
\[
\bZ = \begin{pmatrix}Z_{p+1}\\ \cdots \\ Z_{T+p}\end{pmatrix}
\]
\[ \bep=\begin{pmatrix}\epsilon_{p+1}\\ \vdots \\ \epsilon_{T+p}\end{pmatrix} \]
\[ \bep_* = \begin{pmatrix} \epsilon_{p-q+1}\\\cdots\\\epsilon_p\end{pmatrix} \]
\[
\bhep=\begin{pmatrix}\epsilon_{p-q+1}\\ \cdots \\ \epsilon_1 \\ \cdots \\ \epsilon_{T+p}  \end{pmatrix} = \begin{pmatrix}\bep_*\\ \bep \end{pmatrix}
\]
$$\bhTheta = [\ThetaSTq, \ThetaT]$$
Then the equation (\cref{eq:zeq}) gives:
\begin{equation}
\bZ =  \ThetaSTq  \bep_* + \ThetaT \bep = \bhTheta \bhep
\end{equation}
Hence $v(\bZ)$ is Gaussian:
$$v(\bZ) = (\bhTheta \otimes I_k) v(\bhep)$$
\begin{equation}
\label{eq:covZ}
\Cov(v(\bZ)) = (\bhTheta \otimes I_k)(I_{T+q} \otimes \bOmega) (\bhTheta' \otimes I_k) = \bhTheta \bhTheta'\otimes \bOmega
\end{equation}
$$\Sigma_T = \bhTheta \bhTheta'= \ThetaT\ThetaT' +\ThetaSTq\ThetaSTq' = \ThetaT(I_T + \blambda\blambda')\ThetaT'$$
$$\Cov(\ThetaT^{-1}\bZ) = (I_T+\blambda\blambda')\otimes \bOmega$$
By the Woodbury matrix identity, we obtain the formula for $\Sigma_T^{-1}$:
$$\Sigma_T^{-1}=\ThetaT^{-1\prime}(I_T - \blambda(I_q +\blambda'\blambda)^{-1}\blambda')\Theta_T^{-1}$$
The likelihood formula follows from the fact $\ThetaT^{-1}\bZ$ is Gaussian with distribution $N(0, I_T +\lambda\lambda')\otimes\bOmega)$ with sample values
$$\ThetaT^{-1}\bZ = \ThetaT^{-1}(\bX-\bone \mu -\sum_{i=1^p} L^i\bX\Phi_i)$$
Here we have use the crucial observation that $\btheta(L)$ commute with $\bPhi(L)$ because the former is scalar. The optimal values for $\mu$ and $\Phi$ for a fixed $\theta$ follows from usual GLS analysis.

\section{Implementation}
For small values of $q$, $\ThetaT^{-1}\bX$ could be computed easily via back-substitution at a cost of $Tq$. The rest of the computations are straight forward. Multiplication by $\ThetaT^{-1}$ is the same as convolution with $\btheta(L)^{-1}$, truncated after $T$ steps, so multiplying by $\partial_i\ThetaT^{-1}$ is simply a convolution by $L^i\btheta(L)^{-2}$. This simplifies the gradient calculation to a number of straight forward steps involving matrix multiplication as well as solving small matrix equations. While we do not show all the steps to compute the gradients here, it is available in the open source code.

We have a plan to extend the model to the case where $\btheta(L)$ is a power series to deal with long memory process. For that case, $q$ is infinite but $\btheta(L)$ is dependent on a finite number of parameters. Depending on the data set, it may be useful to explore Fast Fourier Transform. We will not pursue this discussion here.

Let us now discuss the search domain. As mentioned, we will restrict ourselves to invertible $\btheta(L)$. The invertible domain is described by various criteria for stable polynomials. We mention the Schur-Cohn condition as well as the Bistritz tests (\parencite{Schur}, \parencite{Cohn}, \parencite{jury}, \parencite{Bistritz}).

For $q=1$, the domain is simply $-1 < \theta < 1$.

For $q=2$ the domain is given by the inequalities.
\[\begin{gathered}\theta_2 < 1 \\
-\theta_1+\theta_2 +1 \geq 0 \\
 \theta_1+\theta_2 +1 \geq 0\end{gathered} \]
which form a triangle with (inverse) base $\theta_2=1$ and three vertices $(-2, 1)$, $(2, 1)$, $(0,-1)$.

For $q=3$ the conditions are
\[ 1+\theta_1 +\theta_2 +\theta_3 > 0\]
\[ 3+\theta_1 -\theta_2  -3 \theta_3 > 0\]
\[1-\theta_1+\theta_2-\theta_3 > 0 \]
\[ 1-\theta_2 -\theta_3^2 +\theta_1\theta_3 > 0\]
The stability domain is not convex even for $q=3$, see for example \parencite{Ackermann_Barmish}.
We will mostly focus on small $q$ in our examples. The package provides an estimator which can be initialized with the data matrix $\bX$ and the autoregressive degree $p$ as well as specifying if $\mu$ is included or not. For each value of $\theta$ the model computes the negative of the log-likelihood function $LLK(\btheta)$, where the AR polynomials are computed by GLS as specified in the theorem. We provide a fit function for a given $q$, to maximize the likelihood. Fitting is currently done for $q\leq3$ via the constrained trust-region optimization, for higher $q$ we simply assign a very large number for the negative log likelihood (NLLK) to send $\theta$ back to the stability domain. As the stability domain is not convex (however we noted before that the process is well-defined even for non-invertible theta), care should be taken when estimate for $q > 2$. We may need to pick different initial points to optimize globally. For $\btheta$ of first or second order, we provide functions to generate a grid of likelihood function so users can plot and examine the likelihood graphically. The code is available in \parencite{VARsMA_code}. The colab notebook in that repository is available to run interactively.

\section{Simulation studies}
In the first example we take $T=5000, k=2, p=2, q=2$. We use random number generators to generate stable matrices $\bPhi(L)$, $\btheta$ and positive definite $\Omega$. We represent $\Phi$ by a matrix of size $k\times pk = [\Phi_1\cdots \Phi_k]$.

$$\bOmega=\begin{pmatrix} 0.54995831 & 1.15162799 \\
  1.15162799 &  22.99279234 \end{pmatrix}$$
$$\btheta=[0.02992109, -0.55845733]$$
$$ \Phi=\begin{pmatrix} 1.04962255 & -1.45646867 & -0.25126899 & 0.92767515\\
 -0.06188243 & -0.04320034 & 0.03851439 & 0.47572806\end{pmatrix}$$
$$\mu=[1.13078092, 0.10031679]$$
If in the estimator we set $p=2$ with trend, initialize $\btheta$ with a random stable polynomial, the fit function returns:
$$\bOmega_{fit} = \begin{pmatrix} 0.54723465 & 1.08521299 \\
1.08521299 & 22.64522595\end{pmatrix}$$

$$\btheta_{fit} = [-0.04621474, -0.59184145]$$
$$\Phi_{fit} =\begin{pmatrix}  1.12795809 &-0.14321346\\
 -1.45976108 &  0.04203121\\
 -0.2860108 &   0.07812525 \\
 1.04183175 & 0.38799925\end{pmatrix}$$
$$\mu_{fit} = [0.88392291, 0.32717999]$$
$$\nabla \NLLK_{fit} = [4.35765607e-04, 3.21189804e-05]$$
The last expression is the gradient of the negative log-likelhood (NLLK) function at the optimal $\theta$. For the given data $X$ we plot NLLK as a function of $\btheta$ when choosing $q=1$ or $2$ respectively. We see the function is convex in this case, and original parameters of the model is recovered. We also do an extensive test with different choices of $k, p, q$ then regress the coefficients of the data generation process against the fit parameters. Overall, we recover $\bOmega$, while for a regression of coefficients for large $p$ and $k$ does not work quite well. This is probably because of the dynamics between the coefficients which we hope to study further.
\begin{figure}[h]
\includegraphics[scale=0.4]{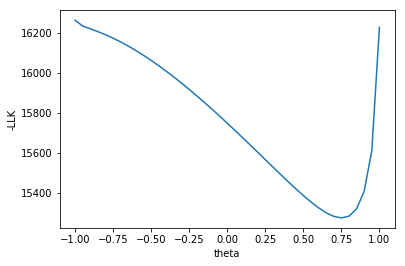}
\includegraphics[scale=0.4]{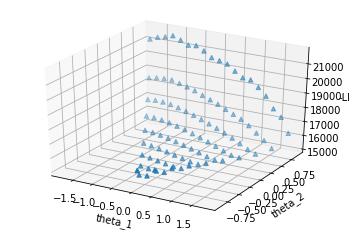}
\caption{LLK for q=1 and 2}
\end{figure}

\begin{figure}[h]
\includegraphics[width=6cm]{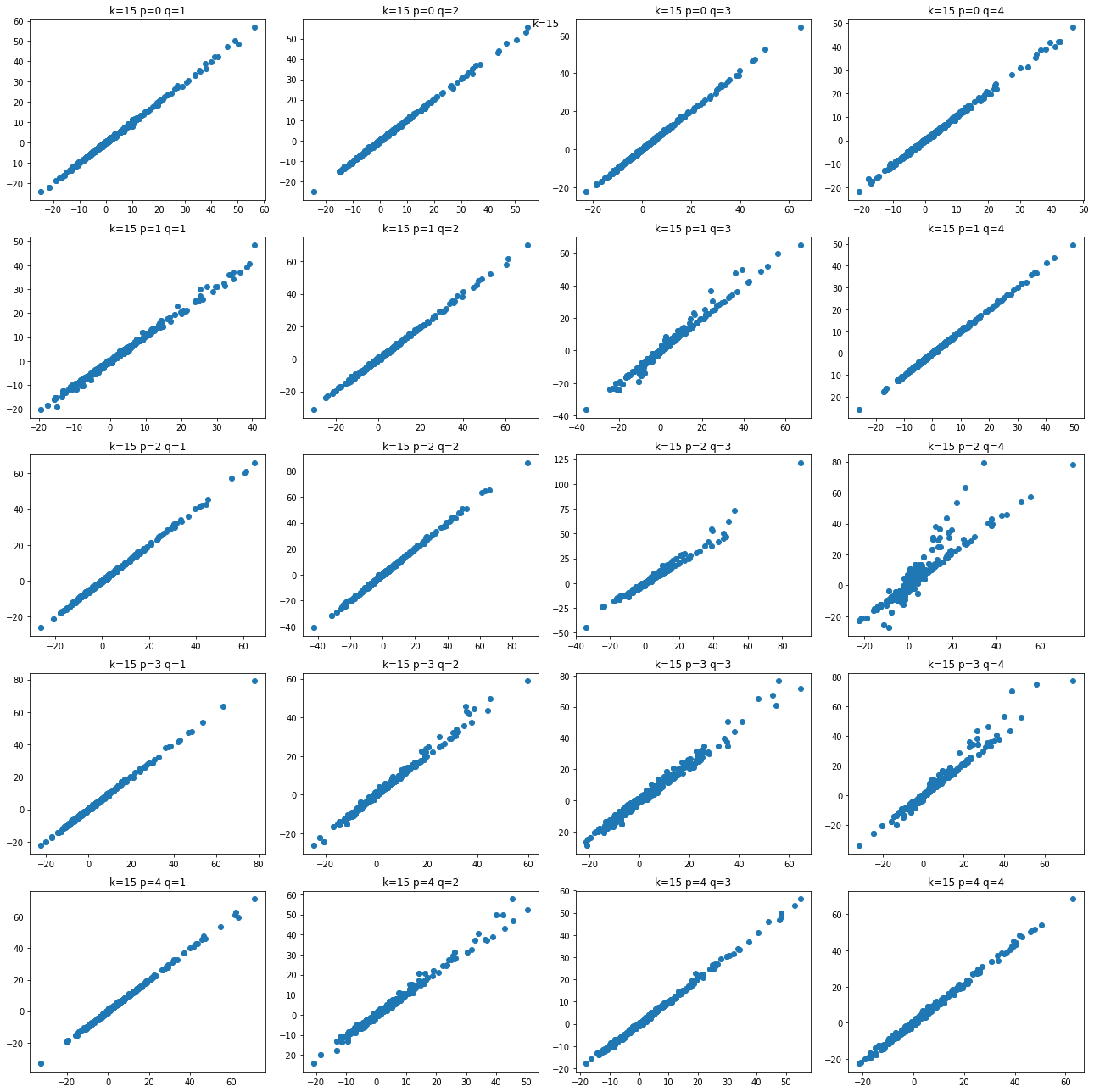}
\includegraphics[width=6cm]{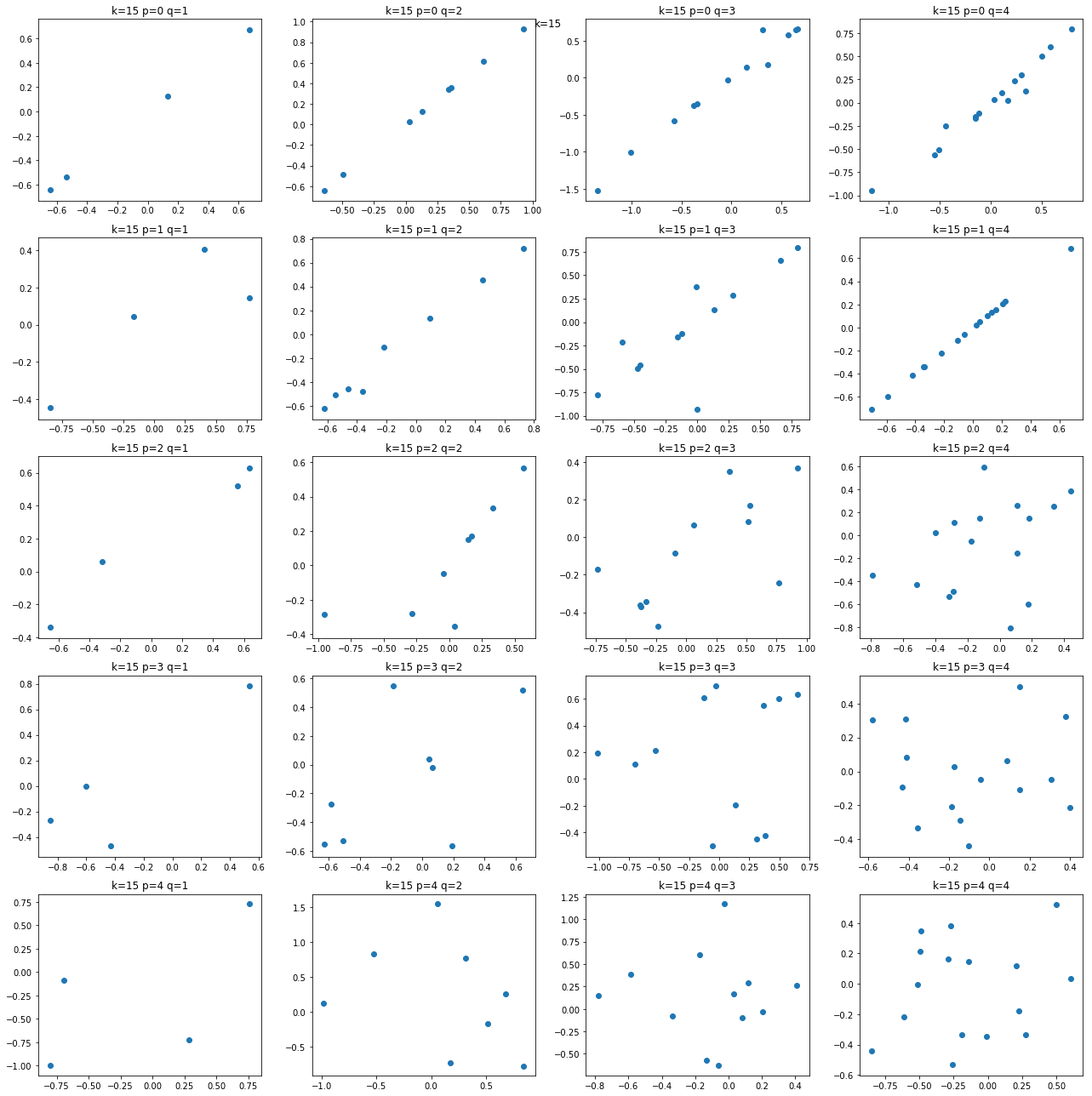}
\centering
\caption{Original v.s. fitted $\bOmega$ (left) and $\btheta$ (right)}
\end{figure}
\section{Relation to Borodin-Okounkov formula and related literature}
\label{sec:history}
We can understand the equations linking $\brSigma_T$ and $\brTheta_T$  as an evaluation of the concentrated covariance matrix $\brSigma_T$ (the left-hand side) by Bayesian theorem. The right-hand side is the result taking expectation over the pre-sample variables $\bep_*$. We note \cref{eq:covZ} in our proof simplified this calculation. Although we reproved \cref{prop:gammaf} only for scalar $\btheta$, the same proof would work for block matrix $\bbrtheta$. It clarifies \cref{prop:gammaf}, as \cite{PhadkeKedem} proved the theorem by inspection. It also gives a probabilistic context to the Borodin-Okounkov's formula.

If the MA component is a power series instead of a polynomial, $\bep_*$ will have infinite dimension. Toeplitz matrices would have to be replaced with Toeplitz operators. Symmetric Toeplitz matrix is also studied in the context of probability and mathematical physics. The Borodin-Okounkov formula is essentially the result $\det(\Sigma_{T\to\infty}) = \det(I + \lambda_{T\to\infty}\lambda_{T\to\infty}')$. In various proofs and extensions of that result over the years, a number of operator identities has been discovered and could be considered as an extension of Phadke and Kedem's identity here.

For example the operator $A$ in the second proof of Borodin-Okounkov formula in \parencite{BasorWidom} could be considered as a generalization of $K(\btheta, T)^{-1}$. Let us restrict to the scalar case for now. 
Our language, $\phi_+(z) = \btheta(z)$, $\phi_-(z) = \btheta(z^{-1})$, $\phi(z) = \btheta(z)\btheta(z^{-1}) = \bgamma(z)$. If $a$ is a Laurent series then the Toeplitz operator $T(a)$ is the matrix with $T(a)_{ij} = a_{i-j}$. So $\bTheta_T$ is just a truncated $T(\phi_+)$, $\bTheta_T^{\prime}$ is a truncated $T(\phi-)$ and $\Sigma_T$ is a truncated $T(\phi) = T(\bgamma)$. Hence
$$A = T(\phi_+^{-1})T(\phi)T(\phi_-^{-1})$$
is related to $\bK(\bTheta, T)^{-1}$.

Given this, one expects it is possible to construct GLS with respect to $\Sigma_T$ constructed from $\btheta(L)$ which is an analytic function depending on a few parameters as opposed to a polynomial (for example a fractionally integrated process). While the analysis may be harder, the modification to the algorithm would be rather straight-forward.

\section{Further directions}
\label{secInt}
From first inspection, the method also could work with seasonality adjustments as well as cointegration analysis. First we note the whole process works if we add additional drift terms, or additional regressions. For example to allow a polynomial drift we add vectors of form $i^k$ instead of $1$ in the definition of $\bX_{\lag}$. Seasonality could be accounted for by seasonal dummy variables, just like the VAR case. 
We will next discuss integrated models. Consider the following model with scalar $\btheta$:
\[ \Phi(L) X = \btheta(L) \epsilon
\]
We note the polynomial division algorithm works for any matrix polynomial and a scalar polynomial. In particular, apply polynomial division of $\Phi$ to $L(L-1)$, note that the remainder matrix is a matrix polynomial of degree at most one we have
\[
\Phi(L) = L(L-1)\Gamma(L)t + \Phi_b(1-L) -\Pi L
\]
(The remainder the division by $L(L-1)$ is of form $A+BL$. we set $\Phi_b = A$, $\Pi = -A-B$.) Let $L=0$ and $L=1$, respectively we get:
\[
\Phi_b = I_k
\]
\[
\Pi = - \Phi_L(1)  = -I_k +\Phi_1+\cdots+\Phi_p
\]
Let $\Delta = 1-L$. The equation becomes:
\[
\Delta \bX(t) =  \Gamma(L)\Delta \bX(t-1) +\Pi X(t-1)+\btheta(L)\epsilon(t )
\]

Apply $\btheta(L)^{-1}$ to both sides we get 
\[
\Delta \bX_{\theta,t} =  \Gamma(L)\Delta L\bX_{\theta,t}  +\Pi L\bX_{\theta, t}+\epsilon_t
\]
where $X_{\theta,t}$ is $\theta(L)^{-1}X(t)$. This is our VECM form. To complete the cointegration analysis we would need a reduced rank version of GLS, which we hope to come back in the future.

\section{Conclusion}
With GLS, we expect many results related to Vector Auto Regressive models are to have corresponding VARsMA analogues. It remains to be seen how the estimation algorithm suggested here applies in practical forecast.
It will need to involve a search for most appropriate values of $p$ and $q$ by using an information criteria, where $q=0$ is the VAR case.

ACKNOLEDGEMENT.
The author is grateful to all who pointed out related works and mistakes and improvements in an earlier version of this paper.
\printbibliography[title={Bibilography}]

\end{document}